# Outside-in stellar formation in the spiral galaxy M33?

F. Robles-Valdez[1*], L. Carigi[1], and M. Peimbert[1]

[1]*Instituto de Astronomía Universidad Nacional Autónoma de México, AP 70-264, 04510 México DF, México*

* E-mail: frobles@astro.unam.mx

**ABSTRACT**

We present and discuss results from chemical evolution models for M33. For our models we adopt a galactic formation with an inside-out scenario. The models are built to reproduce three observational constraints of the M33 disk: the radial distributions of the total baryonic mass, the gas mass, and the O/H abundance. From observations, we find that the total baryonic mass profile in M33 has a double exponential behavior, decreasing exponentially for $r \leq 6$ kpc, and increasing lightly for $r > 6$ kpc due to the increase of the gas mass surface density. To adopt a concordant set of stellar and H II regions O/H values, we had to correct the latter for the effect of temperature variations and O dust depletion. Our best model shows a good agreement with the observed radial distributions of: the SFR, the stellar mass, C/H, N/H, Ne/H, Mg/H, Si/H, P/H, S/H, Ar/H, Fe/H, and Z. According to our model, the star formation efficiency is constant in time and space for $r \leq 6$ kpc, but the SFR efficiency decreases with time and galactocentric distance for $r > 6$ kpc. The reduction of the SFR efficiency occurs earlier at higher $r$. While the galaxy follows the inside-out formation scenario for all $r$, the stars follow the inside-out scenario only up to $r = 6$ kpc, but for $r > 6$ kpc the stars follow an outside-in formation. The stellar formation histories inferred for each $r$ imply that the average age of the stars for $r > 6$ increases with $r$.

## 1 INTRODUCTION

The Triangulum galaxy, or M33, is the third spiral galaxy in size within the Local Group, after the galaxy of Andromeda and the Milky Way. M33 is an Sc late-type galaxy of low-luminosity, about 20 times lower than that of the Milky Way. It does not present a bulge or a bar (Regan & Vogel 1994), and it is located at a distance of 840 kpc (Freedman et al. 1991). M33 has a deprojected semi-major axis of $\sim 9$ kpc (Gil de Paz et al. 2007) and its baryonic mass is about one tenth of that of the Milky Way (Corbelli 2003).

It is important to study M33 because it is intermediate in mass between those galaxies in the Local Group that can be explained as formed by the inside-out scenario (M31 and the MW), and the small mass galaxies of the Local Group that are formed in an outside-in scenario (Hidalgo et al. 2011, Monelli et al. 2011).

The objective of this work is to study the evolution of M33 using chemical evolution models (CEMs) to infer the history of the chemical abundances of the gas content in the disk. We have recent observational data of the surface density of Mgas (Gratier et al. 2010), which shows a double exponential profile: a decreasing one for $r \leq 6$ with and a slightly increasing one for $r > 6$ kpc. We present more reliable values of chemical abundances, values consistent with H II regions and stars. Thanks to the recent increase in the quality of the data, we can infer a reliable chemical history of M33.

There are many models of M33 present in the literature, but they have not been able to explain all the relevant observational data. In particular we want to fit the properties of the central part of the inner disk, and of the outer disk simultaneously.

Throughout this paper we adopt the usual notation *X*, *Y*, and *Z* to represent the hydrogen, helium, and heavy elements abundances by mass, while the abundances by number are represented without italics. In §2 we describe the observational constraints used in this article. In §3 we review previous CEMs of M33. In §4 we present the features of our chemical evolution models. In §5 we present the main results of our models. In §6 we discuss our most interesting outcome within the framework of the Local Group . The conclusions are presented in §7.

## 2 OBSERVATIONAL CONSTRAINTS

In the present work, the chemical evolution models were built to reproduce three main observational constraints of the M33 disk: the radial distributions of the total baryonic mass, the gas mass, and the oxygen abundance, which are shown in Figures 1 and 2. The chemical evolution models were tested with other observational constraints of the M33 disk, such as the radial distributions of: the star formation rate, the stellar mass, the chemical abundances of nine elements and the metallicity, which are shown in Figures 3, 4, and 5. Moreover, the models were built to reproduce the average abundance of iron in the M33 halo, considering the value ‹[*Fe/H*]› = -1.24, estimated for halo stars by Brooks et al. (2004). In this section, we describe the observations used to constrain the models as a function of the galactocentric distance, r, hereafter in kpc units.

### 2.1 Radial distribution of the gas mass surface density, *Mgas (r)*

*Mgas (r)* represents the atomic and molecular gas that contains all the chemical elements in the disk of M33: *X*, *Y*, and *Z* ; i.e. *Mgas (r) = $M_X$ (r) + $M_Y$ (r) + $M_Z$ (r)*. In Figure 1 we show the *Mgas (r)*, as filled black circles, which includes the atomic and molecular components of *X*, *Y* , and *Z*.

To obtain $M_X$ *(r)* we use the updated surface density of hydrogen azimuthally averaged and corrected by the inclination angle by Gratier et al. (2010) from 0.5 to 8.5 kpc.

We added atomic and molecular hydrogen, the sum of both components shows an increase at r > 6 kpc, consequently $M_X$ *(r)* presents a double exponential profile (see Figure 8 by Gratier et al. 2010, and Figure 4 by Verley et al. 2009). In this context, Corbelli & Schneider (1997) observed the distribution of HI in the M33 disc, and found that the gas disk is deformed for *r* > 7.5 kpc, due to warp effects.

Since the helium mass fraction and heavy elements content were not included in the data by Gratier et al. 2010 (private communication, Gratier 2010), we computed $M_Y$ *(r)* and $M_Z$ *(r)* from the following relations:

i) The definition of abundance of an element by mass. Specifically: $M_Y$ *(r)*/$M_X$ *(r) = Y (r)/X(r)* and $M_Z$ *(r)*/$M_X$ *(r) = Z(r)/X(r)*.
ii) The normalization of the chemical abundances by mass: *X(r) = 1 − Y (r) − Z(r)*.
iii) The Y (r) and O(r) enrichment by Carigi & Peimbert (2008): *Y (r) = 0.2477 + 3.3O(r)*.
iv) The transformation of *O/X* (by mass) to O/H (by number): *O(r)/X(r) = 16 (O/H)(r)*.
v) The O/H gradient determined in this paper (see eq. 8): (O/H)(r)= $10^{-0.038r-3.253}$ .
vi) The Z value in the interstellar medium (ISM) is proportional to the O/H ratio, like in the Sun: *Z(r)/(O/H)(r) = $Z_\odot$ /(O/H)$_\odot$* .
vii) The protosolar values of $Z_\odot$ and 12 + log(O/H)$_\odot$ by Asplund et al. (2009), which are 0.0142 and 8.73 dex, respectively.
The contributions of *Y* and *Z* increase the gas mass surface density by a factor of 1.3 approximately, relative to *X*.

### 2.2 Radial distribution of the total stellar mass surface density, *Mstars (r)*

We show in Figure 1, as filled red triangles, the radial distribution of the total stellar mass used to restrict our models.

We took the radial distribution of mass surface density of living stars, *MstarL (r)*, from the profile of the stellar disk surface density from Figure 7 by Corbelli (2003), given by the expression:

$$MstarL\ (r) = 405\ e^{(-r/1.28kpc)}\ . \qquad (1)$$

Recently, Saburova & Zasov (2012) determined a radial profile of the mass surface density of living stars, which is very similar to that by Corbelli (2003). In addition, we took into account the contribution to the stellar mass due to the stellar remnants, $M_{starR}$ *(r)*, with a value equal to the 13% of the living stellar mass, according to that computed in the Galaxy by Carigi & Peimbert (2011), thus:

$$Mstars\ (r) = MstarL\ (r) + MstarR\ (r) = 458\ e^{(-r/1.28kpc)}\ . \qquad (2)$$

The observational profile present a single exponential profile.

## 2.3 Radial distribution of the total baryonic mass surface density, *Mtot (r)*

The total baryonic mass, *Mtot (r)*, shown in Figure 1, as filled cyan squares, is the sum of the gas mass and the stellar mass: *Mtot (r) = Mgas (r) + Mstars (r)*. The radial distribution of the total mass shows a double exponential profile: *Mtot (r)* decays exponentially for $r \leq 6$ kpc and increases slightly for $r > 6$ kpc. Thus, the total mass of the M33 disk is given by:

$$Mtot(r) = \begin{cases} 460\ e^{(-r/1.56\text{kpc})}, & \text{for } r \leq 6 \text{ kpc.} \\ 9.33\ e^{(+r/125\text{kpc})}, & \text{for } r > 6 \text{ kpc.} \end{cases} \qquad (3)$$

Error bars of the data are smaller than the size of the symbols, consequently the behavior of the profile is statistically reliable.

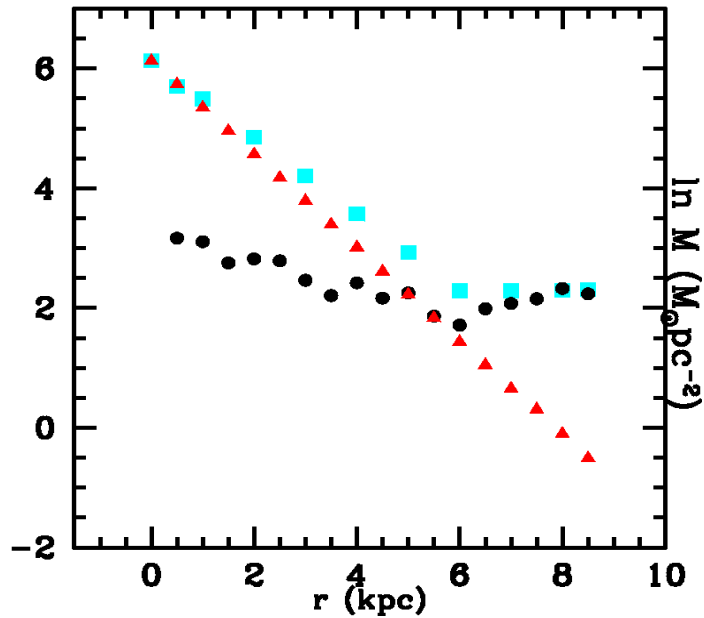

Figure 1. Radial distributions at the present time of: a) the gas mass (filled black circles): atomic and molecular hydrogen by Gratier et al. (2010) including the helium and heavy elements contribution, b) the total stellar mass (filled red triangles): living stellar mass by Corbelli (2003) plus stellar remnants mass according to Carigi & Peimbert (2011), c) the total baryonic mass (filled cyan squares). See sections 2.1, 2.2, and 2.3. The error bars are smaller than the size of symbols.

## 2.4 Radial distribution of the star formation rate, *SFR(r)*

M33 is a galaxy with an active star formation at the present time, as is shown by the large sample of H II regions and O-B stars.

In order to constrain the chemical evolution models, we considered the radial distributions of the star formation rate obtained by Verley et al. (2009). They calculate the transformation of $H_\alpha$ and far ultraviolet (*FUV*) luminosities, which have been corrected by dust extinction, using a stellar population synthesis model with continuous star formation, and they obtain that:

$$SFR(H_\alpha)[M\odot\ yr^{-1}] = 8.3 \times 10^{-42}\ L(H_\alpha)[erg\ s^{-1}]) \qquad (4)$$

$$SFR(FUV)[M\odot\ yr^{-1}] = 8.8 \times 10^{-44}\ L(FUV)[erg\ s^{-1}]). \qquad (5)$$

Verley et al. (2009) also computed the conversion of far infrared luminosity (*FIR*) to *SFR*. The *FIR* luminosity is less affected by dust extinction than the $H_\alpha$ and *FUV* luminosities, however the amount of re-radiation and the stars that heat the dusty environment have to be taken into account. Verley et al. choose the extinction corrected *FUV* as a reference and derive the conversion factor for the *FIR* waveband, obtaining:

$$SFR(FIR)[M\odot\ yr^{-1}] = 13 \times 10^{-44}\ L(FIR)[erg\ s^{-1}]). \qquad (6)$$

## 2.5 Chemical abundances

There are several studies on the chemical abundances of H II regions and O-B stars in M33, which well represent the present-time chemistry in the gas. We considered as observational constraint the O/H values from: H II regions by several authors (see below) and B supergiants by Urbaneja et al. (2005). The data used to built the chemical evolution models in this work have been corrected by the $t^2$ factor and dust depletion. Also we tested our models with other chemical abundances.

Rosolowsky & Simon (2008), using collisional excitation lines (CELs), obtained the O/H abundances in 61 H II regions. Since CELs are sensitive to temperature changes in the H II regions, these O/H values were corrected using the discrepancy factor, known as the $t^2$ factor (Peimbert et al. 2007). It is also necessary to add the dust-phase to the gas-phase component of the O/H values, due to the amount of oxygen atoms trapped in dust grains (Peimbert & Peimbert 2010, PP10). Therefore, O/H values by Rosolowsky & Simon (2008) were increased by 0.23 dex (Esteban et al. 2009, E09) and 0.11 dex (PP10) due to temperature variations and oxygen embedded in dust grains, respectively.

E09, using recombination lines (RLs) obtained the C/H, N/H, O/H, Ne/H, Ar/H and Fe/H abundances in two H II regions: NGC 595 and NGC 604, not included in the sample of Rosolowsky & Simon (2008). The RLs are not sensitive to temperature variations, thus those abundances were not corrected by the $t^2$ factor. We corrected the C/H, O/H, and Fe/H abundances by dust depletion, increasing the gaseous values by 0.10 dex, 0.11 dex, and 1.00 dex, respectively (PP10, E09, Mesa-Delgado et al. 2009). The N/H abundances by E09 were not increased by dust depletion. Since neon and argon are noble gases they are not expected to be embedded in dust grains.

We also considered the O/H ratios determined by Magrini et al. (2010), in 33 H II regions. They used CELs to obtain these abundances, so their values were increased by 0.23 dex (E09) and 0.11 dex (PP10) by the $t^2$ factor and dust depletion, respectively. In addition, we used the S/H and Ar/H values computed by Magrini et al. (2007b) in 14 H II regions. Since those abundances were determined from CELs, we corrected them by the $t^2$ factor, adding 0.21 dex to the S/H values, and 0.24 dex to Ar/H values (E09). The S/H abundances by Magrini et al. (2007b) were not increased by dust depletion. As we mentioned previously, Ar atoms are not embedded in dust grains.

Recombination lines have been used by Bresolin (2011) to determine the O/H abundances in 25 H II regions, but we consider only two of these regions, since the rest of the sample is common to the sample of Rosolowsky & Simon (2008). The O/H ratios by Bresolin were increased by the $t^2$ factor and by dust depletion for 0.23 dex (E09) and 0.11 dex (PP10), respectively.

We used the Ne/H ratios by Willner & Nelson-Patel (2002) obtained from CELs in 25 H II regions. Therefore we corrected the data by the $t^2$ factor, adding 0.26 dex (E09), but not for dust depletion, due to the Ne is a noble gas.

Lebouteiller et al. (2006) derived the P/H abundance in the H II region NGC 604 from CELs. The P/H value was not corrected by either the $t^2$ factor or dust depletion.

The abundances obtained from O, B, and A stars correspond to the present day abundances in the ISM, since these stars have short lifetimes. We use the O/H, C/H, N/H, Mg/H and Si/H abundances estimated by Urbaneja et al. (2005) in 11 B supergiants. Instead of using the Z data by Urbaneja et al., we computed the Z values from the following equation:

$$Z_{stars}/Z_\odot = (C + N + O + Mg + Si)_{stars}/(C + N + O + Mg + Si)_\odot. \qquad (7)$$

adopting the stellar abundances determined by Urbaneja et al. and the protosolar abundances by Asplund et al. (2009). The stellar values do not require corrections by either the $t^2$ factor, or by dust depletion.

In the top panel of Figure 2 we show the O/H aseous values by: Rosolowsky & Simon (2008), E09, Bresolin (2011), and Magrini et al. (2010), without correction by the $t^2$ factor and dust depletion. In the bottom panel we these O/H gaseous values corrected by the $t^2$ factor and dust depletion. Moreover, we plot the stellar O/H abundances by Urbaneja et al. (2005). It is remarkable that the average of the corrected O/H ratios from H II regions is similar to the O/H ratios from B stars. When the gaseous values are not corrected, the stellar abundances are higher, by ∼ 0.3 dex.

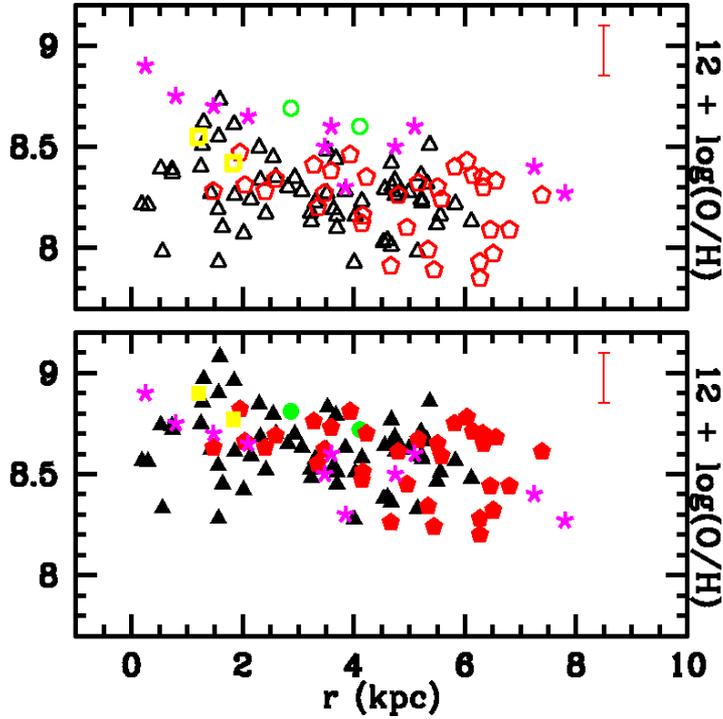

Figure 2. Observed radial distribution at present time of O/H ratios. Top panel: a) gaseous values from H II regions by Rosolowsky & Simon (2008) (empty black triangles), by Esteban et al. (2009) (empty green circles), by Bresolin (2011) (empty yellow squares), and by Magrini et al. (2010) (empty red pentagons), without correction by the $t^2$ factor and dust depletion. b) Stellar values from B supergiants by Urbaneja et al. (2005) (magenta asterisks). Bottom panel: a) gaseous values shown in top panel but now filled symbols, corrected by the $t^2$ factor and dust depletion. b) Stellar values as top panel. See section 2.5. The vertical red line represent the average error.

Considering all the O/H gaseous values corrected by the $t^2$ factor and dust depletion, and the O/H stellar values, we obtain, using least-squares fit, the oxygen gradient as:

$$12 + \log(O/H) = -(0.038 \pm 0.008)\, r + (8.747 \pm 0.035). \qquad (8)$$

We also present abundances for 9 chemical elements and $Z/Z_\odot$ from H II regions (only the corrected ones) and B stars. See section 5.1 for additional details.

Moreover, in order to constrain the efficiency of the star formation rate in the halo, we used the mean abundance of iron, ‹$[Fe/H]$› = -1.24, estimated by Brooks et al. (2004). They obtained $I$ and $V$ photometry of the halo stars of M33, and constructed an $I$ vs. $V - I$ diagram to derive the metallicity distribution function by interpolating between evolutionary tracks of red giant branch models. The value estimated by Brooks et al. is similar to ‹$[Fe/H]$› = -1.27, obtained from halo globular clusters by Sarajedini et al. (2000).

Also we compared the predicted $\Delta Y$ and $\Delta O$ values with those observed in the H II region NGC 604. Then, we considered the He/H ratio obtained by Esteban et al. (2002), and the O/H ratio determined by E09. See section 5.2 for additional details.

## 3 PREVIOUS CHEMICAL EVOLUTION MODELS FOR M33

In this section we discuss some chemical evolution models, CEMs, for M33 published in the literature. Previous models were built to reproduce or compare with: i) gas mass distribution that decays with $r$, even for $r > 6$ kpc and ii) O/H gaseous values without considering the $t^2$ factor and the correction by dust.

Mollá, Ferrini & Díaz (1996), Mollá, Ferrini & Díaz (1997), and Mollá & Díaz (2005) presented a grid of CEMs and compared their results with several galaxies, including M33. In these three articles the CEMs were developed under the SFR prescription by Ferrini et al. (1992). This SFR is a product of the processes of conversion of diffuse gas into clouds and the collapse of these clouds to form stars. Also they used the initial mass function, IMF, by Ferrini et al. (1990), which is similar to that obtained by Kroupa (2001). Mollá and collaborators reproduced reasonably well the O/H abundances in comparing with the observations by Kwitter & Aller (1981) and Vílchez et al. (1988), but they failed ton reproduce the Mgas (r).

Magrini et al. (2007a) used a generalization of the multi-phase model by Ferrini et al. (1992). Their best model assumes an infall rate on the disk almost constant in time, a SFR prescription as Mollá & Díaz (2005) and a two power law IMF, analogous to the IMF by Kroupa, Tout & Gilmore (1993). This model presents a reasonable fit with the O/H abundances from H II regions (Kwitter & Aller 1981, Vílchez et al. 1988, Crockett et al. 2006, and Magrini et al. 2007b). Their predicted Mgas does not agree with the data by Corbelli (2003).

Barker & Sarajedini (2008) built a CEM under the instantaneous recycling approximation. They adopted the IMF by Kroupa, Tout & Gilmore (1993) and used a parametrization of the SFR according to the Kennicutt-Schmidt law (Kennicutt 1998) with a star formation efficiency constant in space and time. They compared the predicted oxygen gradient with the observed one from H II regions and A-B supergiants compiled by Magrini et al. (2007b). They obtained O/H values higher than the observed ones by 0.3 dex. Their predicted Mgas (r) is considerably less than that observed.

Magrini et al. (2010) added to the observational constraints of their previous paper, a new sample of H II regions observed by themselves. Their new CEM is very similar to that built by Magrini et al. (2007a), but now they adopt the IMF by Ferrini, Penco & Palla (1990). Their model continue without reproducing the distribution of gas mass by Corbelli (2003) in the central region (r < 1 kpc) and in the outskirts (r > 6 kpc) of the galactic disk.

Marcon-Uchida, Matteucci & Costa (2010) built a CEM only for the M33 disk based on a version of the MW model by Chiappini, Matteucci & Romano (2001). They adopt the IMF by Kroupa, Tout & Gilmore (1993) and use a Kennicutt-Schmidt SF R with a low star formation efficiency and different thresholds in the gas density for the star formation. The predicted O/H abundances are overestimated for more than 0.25 dex compare to the observed values from H II regions. Moreover for r < 4 their model fails to reproduce the distribution of gas mass by Corbelli (2003) and Boissier, Gil de Paz & Boselli (2007), diverging completely in the inner zone of the disk.

Kang et al. (2012) constructed a CEM under the instantaneous recycling approximation. They used the IMF by Chabrier (2003) and adopted a SF R law correlated with the mass surface density of molecular gas. The predicted O/H fits well the abundances obtained from H II regions, but its absolute value is lower than that determined from B supergiants, since they assume neither dust depletion, nor a $t^2$ factor . They reproduced well the observational gas mass for r < 7 kpc.

Since the previous CEMs cannot explain the main observational constraints for some regions of the disk and nowadays there are more precise observations, we have built new chemical evolution models for M33.

## 4 CHEMICAL EVOLUTION MODELS, CEMS

The CEMs of M33 are based on the standard chemical evolution equations, which were originally written by Tinsley (1980) and are currently used by several authors (e.g. Matteucci 2001, Prantzos 2008, and Pagel 2009). The models have been built to reproduce *Mtot (r), Mgas (r),* and O/H(r) (see section 2). We describe in detail the characteristics of the models below:

i) Halo and disk are projected onto a single disk of negligible width assuming azimuthal symmetry, therefore all functions depend only on the galactocentric distance *r* and time *t*. The single disk is a series of independent concentric rings (1 kpc wide) ranging from 0 to 9 kpc.

ii) M33 was formed in an inside-out scenario with a double-infall of primordial abundances: $Y_p$ = 0.2477 and $Z$ = 0.00 (Peimbert et al. 2007). The adopted double-infall rate is similar to that presented by Chiappini, Matteucci & Gratton (1997), and is given by $IR(r, t) = A(r)e^{-t/\tau_h} + B(r)e^{-(t-1Gyr)/\tau_d}$, where the halo formation occurs during the first Gyr with a timescale $\tau_h$ = 0.5 Gyr (Carigi & Peimbert 2011), and the disk formation takes place from 1 Gyr until 13 Gyr (the age of the model), with a radial increasing timescale $\tau_d = (0.9r + 5)$ Gyr.

The variables $A(r)$ and $B(r)$ are chosen to reproduce the radial distribution of the total baryonic mass at the present-time in the disk component, $M_{tot}(r)$. We also considered that the distributions of disk and halo mass match the disk and halo components of the Solar Neighborhood, $M_{halo}/M_{disk} = 0.25$ (Carigi & Peimbert 2011). Since $M_{tot}(r)$ presents a double exponential profile, this is reflected in the infall rate, taking each one of these variables, $A(r)$ and $B(r)$, two values depending on the galactocentric radius: $A(r) = 92e^{-(r/1.56 kpc)}$ and $B(r) = 368e^{-(r/1.56 kpc)}$ for $r \leq 6$ kpc, and $A(r) = 1.8e^{+(r/125 kpc)}$ and $B(r) = 7.5e^{+(r/125 kpc)}$ for $r > 6$ kpc.

iii) The SFR was parametrized as the Kennicutt-Schmidt Law (Kennicutt 1998): $SFR(r, t) = \nu M_{gas}(r, t)^{1.4}$, where $\nu$ is the star formation efficiency obtained by adjusting the current gas mass radial distribution. In order to match the ‹[$Fe/H$]› shown in halo stars (see section 2.5) we assumed a $\nu$ value 5 times higher during the halo formation than that adopted for the disk.

iv) The models assume the initial mass function by Kroupa, Tout & Gilmore (1993), since the CEMs that considered this IMF reproduce the chemical properties of the Milky Way disk and of the Solar Neighborhood (Carigi & Peimbert 2011). In this work, we considered the mass range from 0.08 M⊙ to an upper mass, $m_{up}$, where $m_{up}$ = 50 M⊙ was chosen to reproduce the current O/H gradient.

v) The models use an array of stellar yields for low and intermediate mass stars, LIMS ($0.08 \leq m/M_\odot \leq 8$), and for massive stars, MS ($8 < m/M_\odot \leq m_{up}$). All these yields depend on the initial metallicity and initial mass of the stars. In order to generate the complete matrix, we interpolated linearly the yields by mass and metallicity, and we extrapolated them, when necessary, by adopting the last available value. The set of yields includes:

a) For stars with masses from 13 to 40 M⊙ and with $Z$ = 0.0, 0.001, 0.004, and 0.02, we adopted the yields by Kobayashi et al. (2006). In the cases of $Z$ = 0.008 and 0.02, we added to the yields by Kobayashi et al. an average of the stellar winds yields of intermediate mass loss rate for He, C, N and O, obtained by Maeder (1992) and Hirshi, Meynet & Maeder (2005), following the suggestion by Carigi & Peimbert (2011). For $Z$ = 0, we increased the yields by Kobayashi et al. adding to them the rotation yields for He, C, N, and O, estimated by Hirshi (2007), that modification was based on the results by Kobayashi, Karakas & Umeda (2011).

b) For star with masses from 1 to 6 M⊙ and with $Z$=0.0001, 0.004, 0.008 and 0.02, we used the yields by Karakas (2010).

c) For binary stars, with total mass between 3 and 16 M⊙, we adopted the yields by Nomoto et al. (1997) in the SNe Ia formulation by Greggio & Renzini (1983). We used the variable $A_{bin}$ = 0.05 as the fraction of binary stars that are progenitors of SNe Ia. This $A_{bin}$ value is chosen in order to reproduce the current Fe/H gradient.

vi) The stellar lifetimes were taken from Romano et al. (2005). These lifetimes are dependent on the initial stellar mass, but independent of initial stellar metallicity.

vii) We do not include radial flows of gas or stars. The loss of gas and stars from the galaxy to the intergalactic medium was not considered.

## 5  RESULTS AND DISCUSSION

We computed two chemical evolution models, CEM1 and CEM2, whose common characteristics were presented in the section 4. These two models are identical for $r < 6$ kpc and the only difference between them is the adopted values of the star formation efficiency, $\nu$, for $r > 6$ kpc.

### 5.1  Radial distributions of Mgas, Mstars, SFR, and Chemical abundances

*5.1.1 CEM with constant ν*

CEM1 was built assuming a constant $\nu$ for $0 < r(kpc) < 9$ and for any time. First, we analyze the CEM1 results for $r \leq 6$ kpc, and then for $r > 6$ kpc.

For r ≤ 6 kpc CEM1 reproduces very well the present time main observational constraints: the *Mgas (r)* and O/H(r) (see Figure 3), according to inside-out galaxy formation scenario. With a given infall, the behavior of *Mgas* depends on the chosen *SFR* and consequently on its ν. To reproduce *Mgas*, we obtain ν = $0.02(M_\odot$ $pc)^{-0.4}$ $Gyr^{-1}$. Then, the oxygen abundance depends on the chosen IMF and its $m_{up}$. To reproduce the absolute value of O/H(r), we obtain $m_{up}$ = $50 M_\odot$.

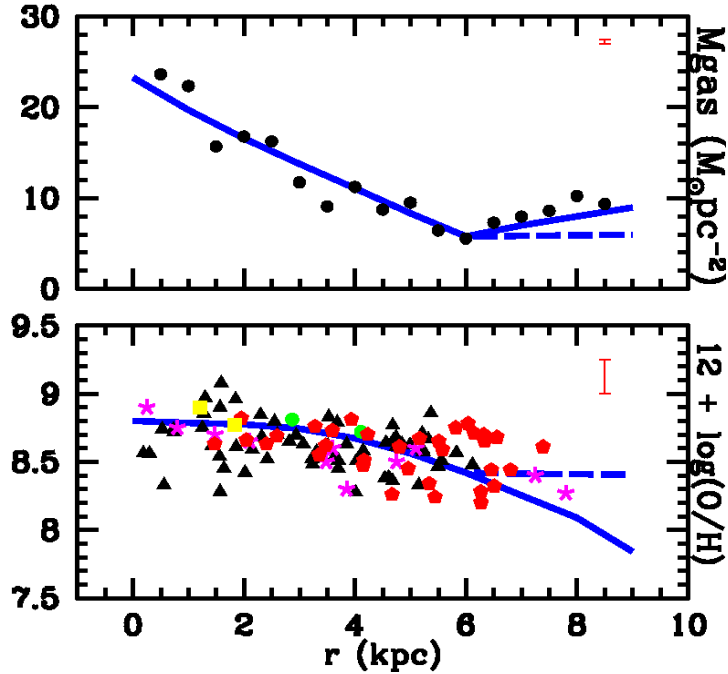

Figure 3. Radial distributions of the gas mass and O/H abundance at the present time, CEM1 (dashed blue line) and CEM2 (best model, solid blue line). Observational data as Figures 1 and 2. The error bars represent the average errors. Note that the gasmass error is smaller than symbol size.

Moreover the predicted *SFR(r)* and *Mstars (r)* show very good fits with observations, see Figure 4. Both decrease with radius, similarly to the *Mgas (r)*, because *Mstars* is a time-accumulative process due to the *SFR* and the *SFR* is proportional to the *Mgas*.

Also the inferred radial distributions of chemical abundances for different elements present a very good agreement with the chemical abundances obtained from H II regions and B stars (see Figure 5), due to the inside-out scenario. In Figure 5 it can be seen that the stellar C observed values are smaller than predicted by the model, while the N observed values are higher than the predicted ones. The increase of N and the decrease of C in the atmospheres of B stars is due to atmospheric build up of N at the expense of C as a consequence of CNO cycled matter in the deepest layers of the convective envelope (e. g. Przybilla et al. 2010, and references therein).

It is important to note that if our model had been built to reproduce the gaseous abundances from CELs without any type of correction, the mup would have been less than 40 $M_\odot$. With this low mup it would be difficult to explain the ionized degree observed in most of the H II regions of M33. Moreover, the corrected gaseous values are in excellent agreement with the stellar abundances, which do not need corrections, see section 2.5 for details.

When we considered the original Ar yields by Kobayashi et al. (2006), the CEMs predicted 0.28 dex lower than those determined in H II regions shown. Therefore the Ar yields were doubled in order to reproduce the absolute value of the Ar/H gradient. The same factor was required by Hernández-Martínez et al. (2011) in the CEMs for NGC 6822, a dwarf irregular galaxy of the Local Group, for matching the Ar/H values in planetary nebulae and H II regions. The Ar yields computed by stellar evolution models could not be checked previously to our work, because the abundance of this element has not been determined from stars in the Solar Neighborhood (Timmes et al. 1995, Romano et al. 2010, Kobayashi et al. 2011).

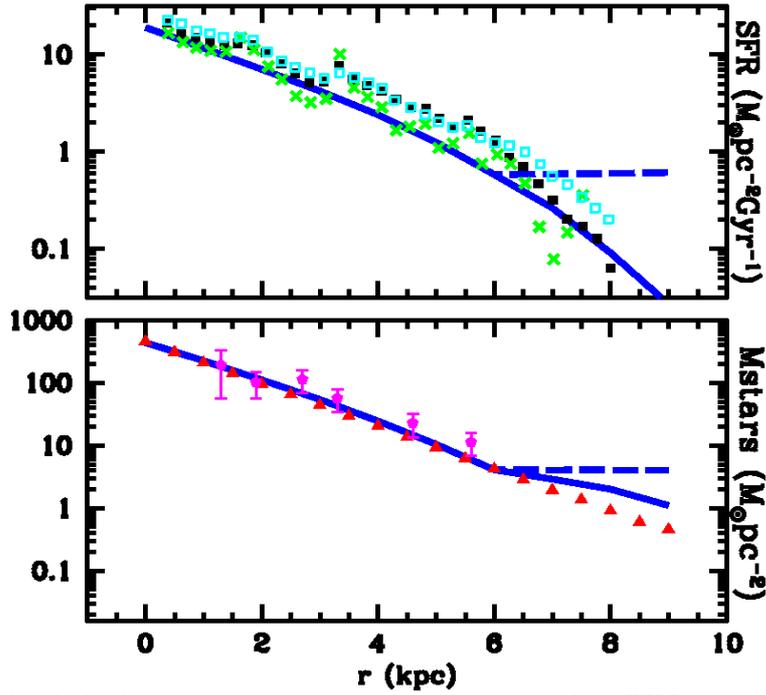

Figure 4. Top panel: radial distribution of the star formation rate at the present time, CEM1 (dashed blue line) and CEM2 (best model, solid blue line). Observational data by Verley et al. (2009): Hα (crosses green), FUV (filled black squares), and FIR (empty cyan squares), see section 2.4. Bottom panel: radial distribution of the stellar mass at the present time, CEM1 (dashed blue line) and CEM2 (best model, solid blue line). Observational data: living stellar mass by Corbelli (2003) and by Saburova & Zasov (2012), plus stellar remnants mass according to Carigi & Peimbert (2011) (filled red triangles and filled magenta pentagons, respectively).

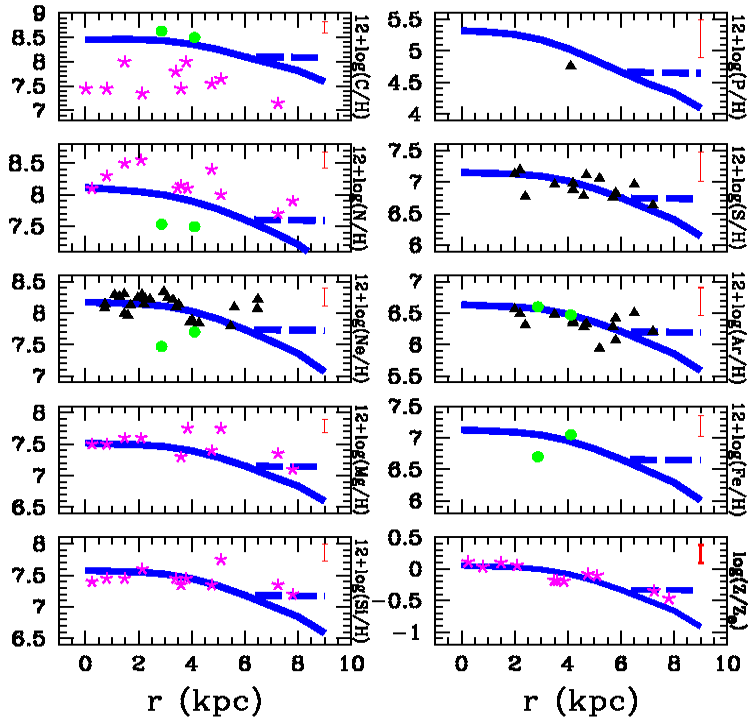

Figure 5. Radial distribution of C/H, N/H, Ne/H, Mg/H, Si/H, P/H, S/H, Ar/H, Fe/H, and Z/Z⊙ abundances, CEM1 (dashed blue line) and CEM2 (best model, solid blue line). Observational data: a) Gaseous and dust values in H II regions by several authors (filled green circles and filled black triangles), b) B supergiants by Urbaneja et al. (2005) (magenta asterisks). See section 2.5. The error bars represent the average errors.

For $r > 6$ kpc CEM1 predicts flattenings for: *Mgas (r)*, *SFR(r)*, *Mstars (r)*, and chemical gradients (see Figures 3, 4, and 5), that is due to the almost flat profile of the total mass adopted, and the slow rate of the gas infall. The predicted *Mgas* is in disagreement with the updated observations, which show an increase of *Mgas* with *r*, increase that might due to gas flows (Spitoni & Matteucci 2011) or to recent mergers (Kewley et al. 2010). That still remains an open problem, because the gas flows are more important towards the galactic center than to the outer regions, and mergers could not explain an azimuthal increase of the *Mgas* surface density. Similarly, the model does not match the *SFR* and *Mstars* as a function of *r*, since the profiles of both observations have a simple exponential decrease with *r*. In the case of the chemical gradients the observational data for the outer disk of M33 are insufficient to test the flattenings predicted by this model.

It is important to note that for r > 6 kpc the observed *Mgas* , *Mstars (r)*, and *SFR* for the present time are not consistent with a Kennicutt-Smichdt law, with constant star formation efficiency. Therefore our CEM1 cannot reproduce simultaneously those observations while we assume a ν constant in space and time.

Therefore for r > 6 kpc, we tried to reproduce the observations by applying in CEM1 different *Mgas* threshold values for the star formation, but the results were not consistent with the observations. Assuming thresholds higher that 5 $M_\odot$ pc$^{-2}$ the *Mgas (r)* was reproduced, but the chemical abundances obtained were extremely low, due to a low star formation and therefore a smaller production of metals. With thresholds lower that 5 $M_\odot$ pc$^{-2}$ the values were very similar to those predicted with CEM1.

*5.1.2 CEM with variable ν. The best model*

Based on the above model, that assumes thresholds in the *SFR*, we realized that the CEMs need a reduction of the *SFR* efficiency only for recent times, but not for past times, in order to explain the current *Mgas (r), Mstars (r), SFR(r)*, and O/H(r) radial distributions for *r* > 6 kpc. If that reduction is considered during most of the evolution, the predicted O/H(r) in the outer regions decreases by several dexs.

Therefore, we built CEM2, with the star formation efficiency, ν, variable in space and time, only for *r* > 6 kpc. We adopted ν = 0.007 ($M_\odot$ pc)$^{-0.4}$ Gyr$^{-1}$ for *r* = 7 kpc, and from *t* = 9 Gyr; ν = 0.002 ($M_\odot$ pc)$^{-0.4}$ Gyr$^{-1}$ for *r* = 8 kpc, and from *t* = 8 Gyr; and ν = 0.0005 ($M_\odot$ pc)$^{-0.4}$ Gyr$^{-1}$ for *r* = 9 kpc, and from *t* = 6 Gyr. That indicates a reduction of *SFR* with the galactocentric distance, earlier at higher *r*, according to the warp effects in the disk (Sánchez Blázquez et al. 2009).

In Figures 3 and 4 we show the very good agreement of *Mgas (r)* and *SFR(r)* obtained from CEM2. Moreover this model considerably improves the agreement of *Mstars (r)* for *r* > 7 kpc, but the predicted values are still 2 times higher than the observations. An updated stellar profile is needed in order to refine the star formation efficiency in the outer part of the disk. In addition, CEM2 predicts chemical gradients for the 10 elements and $Z/Z_\odot$ that adjust very well the observations, see Figures 3 and 5. Based on the improvement in the agreements, we consider that the CEM2 is our best model.

In Figure 6 we present the evolution of the main results of our best model: infall, *Mgas* , *SFR,* and *Mstars* radial distributions at four times: 1, 5, 9, and 13 Gyr. For *r* < 6 kpc, the radial distributions behave as an inside-out model predicts (Naab & Ostriker 2006, Carigi & Peimbert 2008). The high increase in the absolute value of *Mgas* , *SFR,* and *Mstars* between 1 and 5 Gyr is due to the enormous amount of material that falls to form the disk during these 4 Gyrs. At this point it is worth to remind that the disk formation starts at 1 Gyr. For *r* > 6 kpc and *t* = 9 and 13 Gyr the *SFR* decreases with *r* and the *Mgas* increases with *r* due to the level infall rate and the diminished *SFR*.

In Figure 7 we show the evolution of the radial distributions of C/H, O/H, and Fe/H predicted by our best model at four times: 1, 5, 9, and 13 Gyr. For *r* < 6 kpc the gradients evolve as the inside-out scenario sets down (Carigi 1996). For *r* > 6 kpc and *t* ≤ 6 Gyr, the gradients are space flat due to the *Mtot* behavior, the infall, and the *SFR* are almost space constant for those radius, while the absolute values of the gradients increase with time due to the *SFR* increases. Nevertheless, for *t* > 6 Gyr, the gradients get steep due to the reduction of the *SFR* in the outer parts, together with the gas dilution effects of the primordial infall.

In Figure 8 we present the evolution of the star formation rate and infall rate for six galactocentric distances: 2, 4, 6, 7, 8 and 9 kpc. For all radii, the galaxy formation follows a slow inside-out scenario. Specifically, for *r* < 6 kpc, the inner parts were built faster than the outer parts, while for *r* > 6 kpc, the outer parts were built almost simultaneously (bottom panel). Since the *SFR* depends on *Mgas* , *Mgas* depends on infall, and we assumed the same *SFR* efficiency for *r* < 6kpc it follows that the stars formed under an insideout scenario (top panel), as the galaxy formed. Therefore, for $r < 6$ kpc the average stellar age decreases with $r$. Nevertheless, for $r > 6$ kpc we adopted *SFR* efficiencies that decrease earlier with $r$, causing a stellar formation that follows an outside-in scenario (central panel), while the galaxy formation is almost homogeneous. Consequently, for $r > 6$ kpc, the average stellar age increases with $r$.

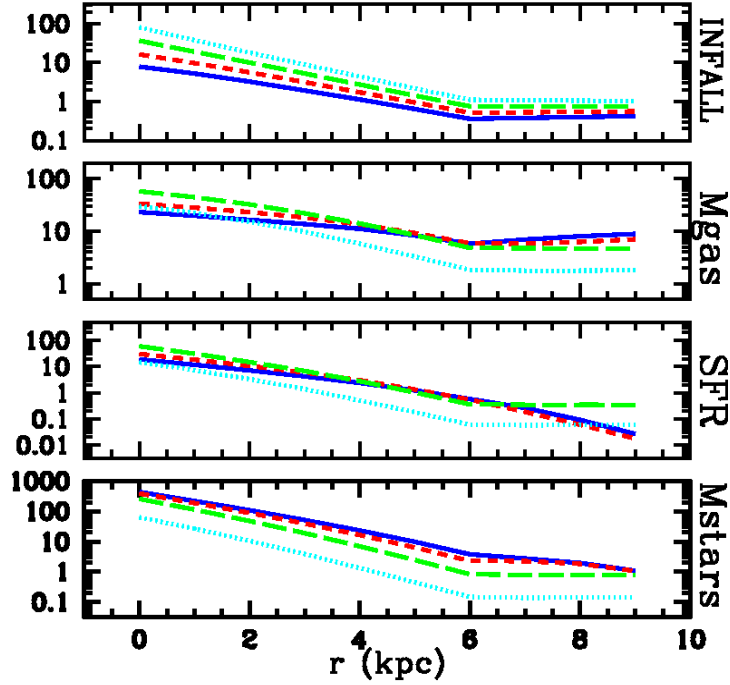

Figure 6. Radial distributions of: Infall ($M_\odot$ pc$^{-2}$ Gyr$^{-1}$), Mgas ($M_\odot$ pc$_{-2}$), SFR ($M_\odot$ pc$^{-2}$ Gyr$^{-1}$), and Mstars ($M_\odot$ pc$^{-2}$) for our best model (CEM2) at four times: 1, 5, 9, and 13 Gyr (dotted cyan, long-dashed green, short-dashed red, and solid blue lines, respectively).

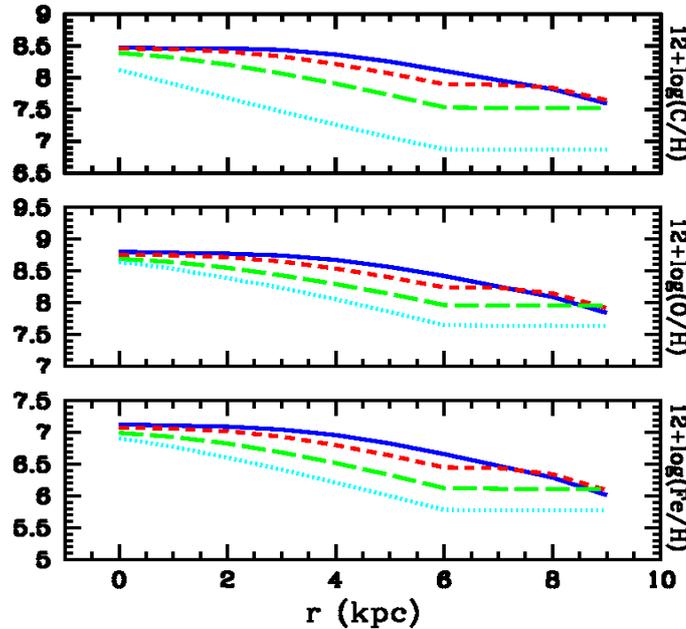

Figure 7. Radial distributions of C/H, O/H, and Fe/H abundances for our best model (CEM2) at four times: 1, 5, 9, and 13 Gyr, lines as Figure 6.

From our best model, we show that the variations in the star formation efficiency for $r > 6$ kpc can match better the *Mgas (r)* and *SF R(r)* observed, the *Mstars (r)* and the chemical gradients. Therefore a lower *SFR* and variable in space and time can explain in a more refined way the behavior for outer radii. The decrease of the *SFR* may be due to the reduction of molecular gas in external radii necessary to form stars, or to warp effects as mentioned previously.

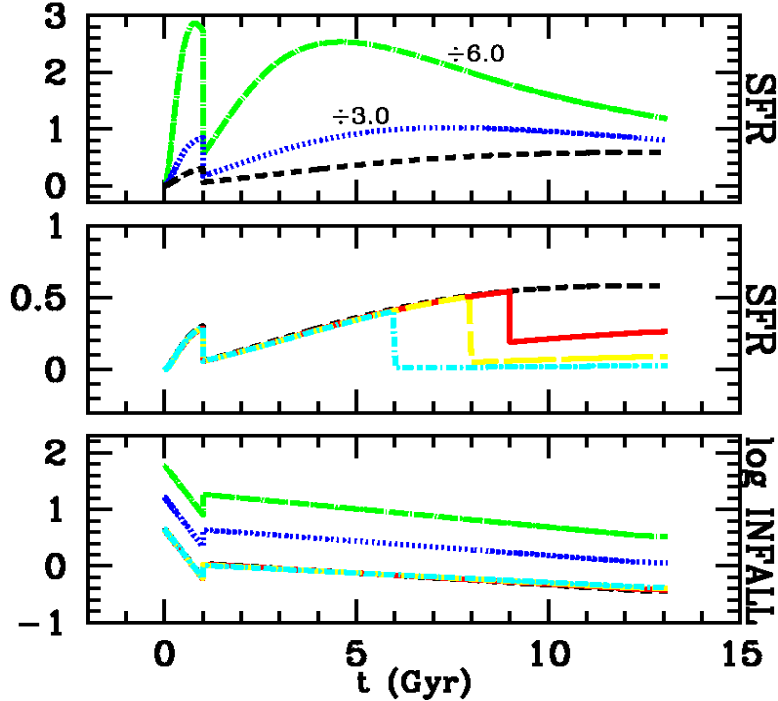

Figure 8. Evolution of: *SFR* ($M_\odot$ pc$^{-2}$ Gyr$^{-1}$), top and medium panels, and of: Infall ($M_\odot$ pc$^{-2}$ Gyr$^{-1}$), bottom panel, for our best model (CEM2) at six radii: 2, 4, 6, 7, 8 and 9 kpc (dot-long-dashed green, dotted blue, short dashed black, solid red, long-dashed yellow, and dot-short-dashed cyan lines, respectively). In the top panel the *SFR* for $r$ = 2 and 4 kpc have been divided by 6 and 3 respectively.

## 5.2 The ΔY /ΔO ratio

The study of the Δ*Y* /Δ*O* ratio is important to check the consistency of the chemical evolution models since He is produced by LIMS and MS, while O is only produced by MS. For this reason the Δ*Y* /Δ*O* ratio depends on: i) the IMF, ii) the SFR because Δ*O* is in agreement with the instant recycling approximation and Δ*Y* is not, and iii) the *Y* and **O** yields because they are metal dependent.

The only H II region in M33 that has determinations of Δ*Y* and Δ*O* of high accuracy is NGC 604 (Esteban et al. 2002, 2009). For the He/H ratio we will use the results by Esteban et al. (2002) that considered the effect of the underlying absorption of the He I lines, while E09 did not. For the O/H ratio we will use the results by E09 because they did observe the 3727 [O II] line and determined with higher accuracy the O II recombination line intensities than Esteban et al.(2002). The gaseous O/H ratio amounts to 12 + log(O/H) = 8.60. The total value including the fraction of O atoms embedded in dust amounts to 8.71 dex (PP10).

To determine the production of helium in M33 we need to compare the present abundance of *Y* with the primordial helium abundance *Yp* . The Y determination by Esteban et al.(2002) is based on the He+ recombination coefficients by Smits (1996) while the primordial helium value of *Yp* = 0.2477 by Peimbert et al. (2007), used in this paper, is based on the He+ recombination coefficients by Porter et al. (2005, 2007). Therefore to determine Δ*Y* = *Y* − *Yp* we recomputed the values by Esteban et al. (2002) using the results by Porter et al. and found that the *Y* value of 0.2641 by Esteban et al. had to be increased to 0.2682, it should be noted that as long as the *Y* and *Yp* abundances are computed with the same He+ recombination coefficients, Δ*Y* is not affected by small improvements on the He+ recombination coefficients.

In Table 1 we present the observed Δ*Y* /Δ*O* for NGC 604 and compare it with that predicted by our chemical evolution model. We have considered a galactocentric distance of 4.11 kpc for NGC 604. We also present the Δ*Y* /Δ*Z* ratio where we have assumed that the *O* abundance represents 43.3% of the total heavy element abundance *Z*, as is the case in the protosolar abundances (Asplund et al. 2009).

In Figure 9 we present the evolution of Δ*Y* versus Δ*O* and of the Δ*Y*/Δ*O* ratio predicted by our model for *r* =4.11 kpc and compare both results with the observed values. Since CEM1 and CEM2 are identical for r ≤ 6 kpc, we present only the CEM2 results. The good agreement between the model and the observational restrictions imply that the ingredients used in the model, mainly the IMF and the yields are adequate for this galaxy. For other galactocentric radii Δ*Y* vs Δ*O* the evolution of M33 is similar to Δ*Y* vs Δ*O* evolution of the Milky Way, see Figure 7 by Carigi & Peimbert (2011).

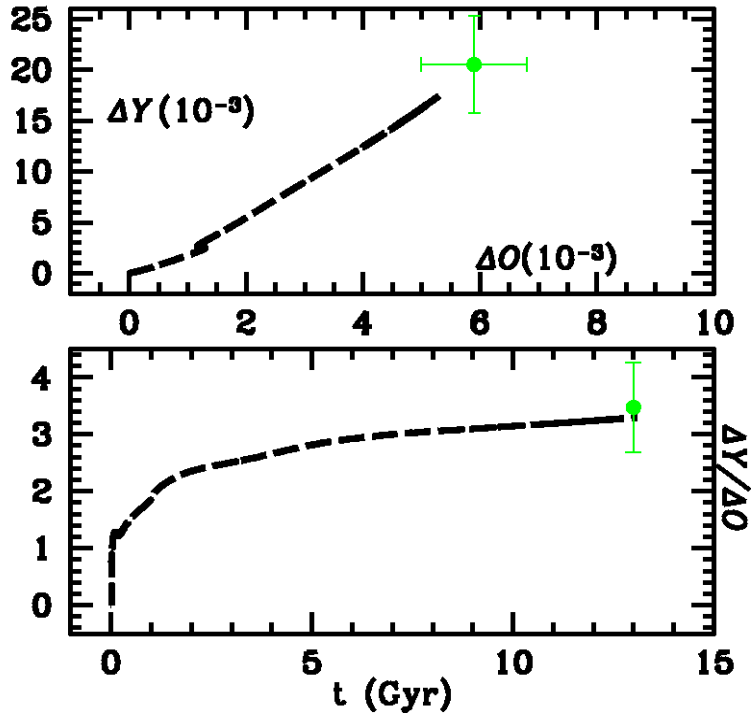

Figure 9. Chemical evolution models at r = 4.11 kpc (long-dashed black). Top panel: Δ*Y* vs Δ*O* evolution. Bottom panel: Δ*Y*/Δ*O* evolution. Observational data: NGC 604 at r = 4.11 kpc by Esteban et al. (2002, 2009) (green filled circles). The shown values include the corrections by the $t^2$ factor (for He and O) and by dust depletion (only for O). See section 5.2 for details.

*Table 1. Abundances of Helium, Oxygen, and Heavy Elements by Mass.*

| Parameter | $t^2 \neq 0.000$ | CEM2 |
|---|---|---|
| Y | $0.2682^a \pm 0.0040$ | 0.2651 |
| ΔY | $0.0205^b \pm 0.0048$ | 0.0174 |
| O | $0.00590^c \pm 0.00090$ | 0.00532 |
| ΔY/ΔO | 3.47 ± 0.79 | 3.29 |
| Z | $0.0139^d \pm 0.0021$ | 0.0116 |
| ΔY/ΔZ | 1.47 ± 0.34 | 1.51 |

a  This value is taken from Esteban et al. (2002).
b  Considering Yp = 0.2477 by Peimbert et al. (2007).
c  This value is taken from Esteban et al. (2009).
d  Considering that the O is the 43% of the value of Z.

## 6 INSIDE-OUT AND OUTSIDE-IN STELLAR FORMATION IN THE LOCAL GROUP OF GALAXIES

The Local Group is formed by two massive spiral galaxies (M31 and the MW), about 70 minor galaxies (Irr, cE, dIrrs, dSphs, dIrr/dSphs) and one small spiral galaxy (M33) (McConnachie 2012).

The spiral galaxies of the Local Group present negative B-V and chemical gradients, indicating a negative age gradient, being the inner parts dominated by old populations and the outer parts by young populations. That suggests an inside-out stellar and galactic formation (Carigi & Peimbert 2011, Carigi, Meneses-Goytia & García-Rojas 2012).

Many dwarf galaxies of the Local Group present positive B-V gradient and consequently positive age gradient, being the inner parts dominated by young populations and the outer parts by old populations. That suggests an outside-in stellar and galactic formation (de Boer et at. 2012a,b; Hidalgo 2012; Hidalgo et al. 2012 and references therein).

In our work, we found that the inner parts of M33 ($r < 6$ kpc) behave like M31 or MW (the mean stellar age decreases with radius), while the outer parts of M33 ($r > 6$ kpc) behave as dwarf galaxies (the mean stellar age increases with radius). Therefore M33, a galaxy with an intermediate mass, presents characteristics similar to the smallest and the biggest galaxies, in the respective mass regimes, being a unique example in the Local Group that connects the formation and evolution of both types of galaxies.

In the outermost disk of some spiral galaxies outside the Local Group, flattening in the chemical gradients has been observed (Vlajic 2010; Bresolin, Kennicutt & Ryan-Weber 2012). The determination of the flatten gradient at high $r$ is still an open question for the MW and M31 (Esteban et al. 2012 and references therein), due mainly to the inconsistency among stellar and gaseous data, and it is not found yet for M33, due to the lack of data (see section 2.5).

Different processes offer possible explanations for the chemical gradient flattening: a) radial stellar migrations (Roediger et al. 2012, Minchev et al. 2012) that provide the outer disk with stars born in the inner part of the galaxy and therefore enrich these external zones, b) effects of the warp that can produce decreases in the star formation (Sánchez-Blázquez et al. 2009), c) physical conditions in the distribution of gas not adequate for the star formation (deficient amount of molecular gas, gas temperature, among others), d) mixing and turbulence processes, (e) galactic scale out-flows, and (f) enriched accretion (Bresolin et al).

It is important to mention that a nearly flat chemical gradient has been estimated in dwarf galaxies, based on stellar Fe/H and gaseous O/H observed values and on theoretical Z values inferred by synthetic Color-Magnitude Diagram analysis (Tolstoy, Hill & Tosi 2009).

It is worthy to remark that the direct association of the accreted baryonic mass, or galactic formation, with the SFR intensity, or stellar formation, (and vice versa) does not follow in the outer parts of M33. Consequently, the simple inference of how a galaxy forms from the star formation history could be risky, because the SFR depends on other factors, in addition to the gas amount.

## 7　CONCLUSIONS

Based on an exhaustive review of the recent observational data of M33 we computed a successfully chemical evolution model for this galaxy. From this model we reach the following conclusions.

1) The total baryonic mass surface density shows a decreasing exponential behavior as a function of $r$ up to $r = 6$ kpc, beyond this radius it shows a light rise, due to the increase of the gaseous mass surface density for $r > 6$ kpc.

2) O/H abundance ratios from O-B stars agree with abundances derived from H II regions only if the RLs gaseous abundances are corrected by dust depletion. Moreover the O/H abundance ratios from O-B stars agree with CEls O/H gaseous abundances only if they are increased by the effect of temperature variations and by the fraction of O atoms embedded in dust grains.

3) From the stellar and corrected gaseous abundances, the oxygen gradient of the disk is: $12 + \log(O/H) = -(0.038 \pm 0.008)r + (8.747 \pm 0.035)$.

4) To reproduce the absolute value of the observed O/H gradient, we obtain that the IMF upper mass for M33 amounts to 50 $M_\odot$.

5) To reproduce the absolute value of the Ar/H gradient, the Ar yields need to be two times higher than the ones computed by Kobayashi et al. (2006).

6) To match the absolute value of the Fe/H gradient, the fraction of binary systems that become SNe Ia has to be 5%. To reproduce the < [F e/H] > shown by the halo stars, the efficiency of star formation in the halo has to be 5 times larger than that adopted for the disk.

7) The $\Delta Y/\Delta O$ value equal to 3.29, predicted by the model for $r = 4.11$ kpc, is in very good agreement with the observed value: 3.47 ±0.79 for NGC 604, an H II region of M33.

8) No star formation thresholds based on the surface density of gas mass could explain the galactic properties for $r > 6$ kpc. It is necessary to determine thresholds as a function of the volumetric density of gas mass for the purpose of understanding the dependence on time and galactocentric radius found for the outskirts of the disk.

9) For $r > 6$ kpc, updated determinations of: stellar mass, star formation, and chemical abundances are required to try to explain the formation and evolution of the M33 outer part.

10) The inner part (r < 6 kpc) of M33 was formed according to a slow inside-out scenario, but the outer part ($r > 6$ kpc) was formed almost independently of the galac- tocentric distance.

11) For the inner part, the star formation efficiency is constant in time and space, but for the outer part the efficiency decreases with time and galactocentric distances. The reduction of the SFR efficiency, earlier at higher $r$, produces an outside-in stellar formation and consequently an increase with $r$ of the average stellar age. That radial-temporal behavior of the SFR efficiency for $r > 6$ kpc may be caused by the warp in the disk. Warps can reduce the gas volumetric density, even when the gas surface density is kept constant or increases.

We thank the referee for a careful review of the manuscript and several useful suggestions. F. Robles-Valdez is grateful to CONACyT, México, for a doctoral grant. L. Carigi thanks the funding provided by the Ministry of Science and Innovation of the Kingdom of Spain (grants AYA2010-16717 and AYA2011-22614). This work was partly supported by grants 60354 and 129753 CONACyT.